# Probability distribution of substituted Titanium in $RT_{12}$ (R = Nd, Sm, T = Fe, Co) structures

C. Skelland[1], T. Ostler[7], S. C. Westmoreland[2], R. F. L. Evans[1], R. W. Chantrell[2], M. Yano[3], T. Shoji[3], A. Manabe[3], A. Kato[3], M. Winklhofer[4], G. Zimanyi[5], J. Fischbacher[6], T. Schrefl[6], and G. Hrkac[1]

[1]College of Engineering, Mathematics and Physical Sciences, University of Exeter, UK
[2]Department of Physics, University of York, York, YO10 5DD, UK
[3]Toyota Motor Corporation, Toyota City, 471-8572, Japan
[4]University of Duisburg, Germany
[5]University of California, UCDavies, USA
[6]Center for Integrated Sensor Systems, Danube University Krems, Viktor Kaplan Straße 2 – E 2700 Wiener Neustadt, Austria
[7]Faculty of Arts, Computing, Engineering and Sciences, Sheffield Hallam University, Howard Street, Sheffield, S1 1WB, UK

**We investigated the atomic fill site probability distributions across supercell structures of $RT_{12-x}Ti$ (R=Nd, Sm, T=Fe, Co). We use a combined molecular dynamics and Boltzmann distribution approach to extrapolate the probability distributions for Ti substitution from lower to higher temperatures with an equilibrium condition to assess how temperature affects the predictability of the structures fill path. It was found that the Nd and Sm based Fe systems have the highest filling probability path at lower temperatures but the cohesive energy change due to Ti substitution in Sm and Nd based crystals indicates that a more stable system could be achieved with a combination Co and Fe in the transition metal site.**

## I. INTRODUCTION

Permanent magnetic materials have been an important area of research for the last century [1], starting with high carbon steels in the early 1900's and moving on to high performance rare earth magnets near the end of the century. As their properties have improved their application has grown and they are now used in a wide variety of products such as printers, hard disks, MRIs and electrical engines. Of these areas of application their use in electrical engines is the most important for the realisation of a future that runs entirely on renewable energy sources. Their use in the engines of electric vehicles and renewable energy technologies such as wind turbines is of utmost importance to these devices efficient performance.

The performance of a high density permanent magnet can be classified by its maximum energy product ($BH_{max}$), that is governed by its remanent magnetisation, and coercivity. As these properties are temperature dependent and energy applications of magnets operate at high temperatures e.g. electrical engines operate at 450-475K, it is important to understand the fundamental morphological and chemical properties that govern this temperature dependency.

The current permanent magnetic material used in electrical engines, $Nd_2Fe_{14}B$, has a curie temperature of 585K as measured by Sagawa et al. [2], but can be improved through substitution of Neodymium with Dysprosium to reach a higher curie temperature, and this is the predominant method currently used to improve the properties of $Nd_2Fe_{14}B$ at high temperatures. However due to the scarcity and therefore prohibitively high cost of Dysprosium, as well as its potentially tenuous supply [3], there is a great deal of pressure to find new permanent magnetic materials made of abundant and hence cheaper constituents that have similar magnetic properties but a higher curie temperature.

One of the most promising of these materials is the $RT_{12}$ (R = rare earth, T = transition metal) phase, which forms in the $ThMn_{12}$ structure. The phases main problem is its intrinsic instability in binary form meaning it must be stabilised with a ternary element to form an $RT_{12-x}M_x$ structure (M = Si, Ti, V, Cr, Mo or W). It was found by De Mooij and Buschow [4] that of the possible phases the $RFe_{12-x}M_x$ phase was the most promising due to the high magnetic moment and anisotropy provided by Iron. Further investigation of this phase with M = Ti by Yang et al. [5] found that nitrogenation of these structures causes significant changes to their magnetocrystalline anisotropy and via this method their magnetic properties could be significantly improved. Of the structures investigated $NdFe_{11}TiN$ was thought to be the most promising with a curie temperature of 740K and estimated $BH_{max}$ of 455kJ/m$^3$, which compares well to $Nd_2Fe_{14}B$'s theoretical $BH_{max}$ of ~500kJ/m$^3$ and curie temperature of 585K [6].

Recent computational investigation of this phase by Miyake et al. [7] has shown that Titanium substitution changes crystal electric field parameter $A_{20}$ from -83K in $NdFe_{12}$ to +54K in $NdFe_{11}Ti$. Nitrogenation of this structure further increases $A_{20}$ to +439K. Titanium substitution whilst stabilising the structure also causes a large loss of spin magnetic moment per formula unit of: -4.90$\mu_B$ from $NdFe_{12}$ to $NdFe_{11}Ti$, and -5.16$\mu_B$ from $NdFe_{12}N$ to $NdFe_{11}TiN$, this is due to it aligning anti parallel to the internal magnetisation and decreasing the magnetic moment of the surrounding Iron atoms. Therefore, it is important that Titanium substitution is kept to a minimum and that its behaviour inside the structure is well understood.

Our work is concerned with the investigation of the $RT_{12-x}Ti_x$ (R = Sm, Nd; T = Fe, Co) phases structural properties and how they change with Titanium substitution. We use molecular dynamics with Morse potentials and Boltzmann statistics to investigate how replacement of the transition element with a tertiary element changes the calculated cohesive energy of the structure at low Titanium at. %, as well as investigating the probability distribution of Titanium over the atom positions within the structure.

## II. METHODOLOGY

In order to calculate the structural properties of the $RT_{12-x}Ti_x$ phases we use molecular dynamics, which uses the classical equations of motion with parameterised force fields to predict the structure of physical systems [8]. As described in our previous work [9] interatomic potentials for metals are used to simulate atom interactions in our investigated system. We use Morse [10] and modified embedded atom models (MEAM), which are derived from ab initio and validated with lattice constants from experiments and literature. Our calculated lattice constants for $SmFe_{11}Ti$, $NdFe_{11}Ti$, and $SmCo_{11}Ti$ were in agreement with those found in the literature ([5], [5], [11] respectively) to within <2% (see Table 1), whilst the Ti replacements followed the expected pattern set out by De Mooij and Buschow [4].

**TABLE 1 HERE**

The metallic phase is modelled with a Morse potential with the form:

$$\Phi_{sr}(r_{ij}) = D_{ij}[(1 - e^{-B_{ij}(r-r_0)})^2 - 1] \quad (1)$$

Where $D_{ij}$ is the disassociation energy of the bond and $\beta_{ij}$ is a variable parameter that can be determined from spectroscopic data, and the potential energy of the system was minimised via the Newton Raphson method in GULP [8].

The set of potentials pertaining to Nd in the $NdFe_{12-x}Ti_x$ simulation are shown graphically in figure 1.

**FIG. 1 HERE**

Using these potentials energy minimisation calculations are performed on the investigated structures at constant pressure until they reach a local energy minima, the structural properties as well as the cohesive energy of this optimised structure are then output, with the minimum energy structure being used to inform the next round of calculations. In this way, Titanium is permutated through the structure up to the desired stoichiometric percentage.

Using the output data and basing on energy we calculate the probability ratios for each position being filled at increasing stoichiometric Ti at. % up to 4%. The ratios are based off of Boltzmann statistics and modelled on the Boltzmann distribution equation [12]:

$$p(r) = \frac{e^{\frac{-E_r}{k_BT}}}{\sum_i e^{\frac{-E_i}{k_BT}}} \quad (2)$$

Where $p(r)$ is the probability of state $r$, $E_r$ is the energy of state $r$, $k_b$ is the Boltzmann constant, and $T$ is the temperature in Kelvin. These probabilities are then normalised by the probability of the minimum energy position, which gives probability ratios which can be calculated by the equation below:

$$\frac{p(state2)}{p(state1)} = e^{\frac{E_1-E_2}{k_BT}} \quad (3)$$

The ratios are further normalised by the summation of all ratios to give each positions percentage probability at a particular Ti at. %. Whilst the simulations were run at 0K it has been assumed that probabilities at higher temperatures can be obtained by scaling with $k_BT$ and making the assumption that these systems have been left to find equilibrium, so that necessarily the most probable position will be filled by the replacing Titanium.

## III. RESULTS

Simulations calculating Ti atom distribution within singular unit cells of the three investigated structures showed clear differences between the probability of the 8i, 8j, and 8f Wycoff positions in the $ThMn_{12}$ phase structure. The 8i position set has the highest probability for replacement followed by the 8j and 8f sets, this is in agreement with the results from T. Miyake [7]. This probability distribution is shown in figure 2, along with a diagram of the 1:12 phase structure noting the three sets of Wycoff positions.

Due to the difference in probability between the position sets at 300K being of the order $10^{19}$ and $10^{32}$ for 8i/8j and 8i/8f respectively it is reasonable to assume that at <4% Ti at. % the only positions of significance will be those in the 8i position set, and this is exactly what was seen by De Mooij and Buschow [4]. All future simulations therefore disregard all but the 8i positions in an effort to improve computational efficiency.

Supercell structures were investigated in order to decrease the Ti percentage in the structure to <4% and down to ~1 Ti at. %. The results of these simulations are given below. The 8i positions within these supercell structures are given in Table II.

**FIG. 2 HERE**

*A. NdFe12:*

Titanium replacements align around a singular Nd atom with the first substitution designating which of the atoms it will be, for example the first position may be 5. The first substitution is followed by a second which goes with equal probability to one of the next two nearest neighbours that surround the Nd atom, positions 21 and 29 in the case of 5 being the first substitution.

This repeats up to ~4 Ti at. % at which point the structure has a layout analogous to that shown in figure 3a. The path taken to the structure pictured is (position) 5, 21, 13, 29.

### B. SmCo12:

The first two Titanium replacements in this unit cell follow the same pattern as those in NdFe$_{12}$ and start by surrounding a Samarium atom, however on the third substitution the Titanium replacement fills an analogous position to the one it would have done in NdFe$_{12}$ but in a different unit cell. The unit cell the replacing atom goes to is dependent on which similar position it would have filled. For example, if position 5 then 21 are filled, the third position could be 14 or 31, which are similar to positions 15 and 29 respectively. The final substitution surrounds a further none similar Samarium atom, and when extending the cell boundaries, it becomes clear that the Titanium has substituted itself such that it forms lines that run through the solid roughly along the [110] or [-110] direction. The direction of the line is dependent on the second atom substitution, and the [110] direction structure at ~4 Ti at. % is shown in figure 3b. The path taken to the structure pictured is (position) 1, 18, 12, 25.

**TABLE 2 HERE**

### C. SmFe12:

The first three Titanium replacements follow the same pattern seen in SmCo$_{12}$ and align themselves around a singular Sm atom before filling a similar position that is analogous to one that would have been filled in NdFe$_{12}$. However, on the final atom, rather than spreading out evenly between the unit cells as in SmCo$_{12}$, the substitution fills the final space surrounding the first Sm which it hasn't already similarly filled elsewhere in the structure. For example, if positions 5, 21, and 14 are filled then the final fill position would be 13. This substitution layout forms a zig zag pattern moving in the [100] or [010] direction, with the direction of the pattern dependent on the third substitution. The final structure in the [100] direction at ~4 Ti at. % is shown in figure 3c. The path to get to the pictured structure is position 1, 18, 27, 11.

The three structures probability distributions predict that they behave the same way at or below ~2 Ti at. %, with divergent distributions occurring at ~3 Ti at. % and above for the Nd and Sm based systems and at ~4 Ti at. % for the Co and Fe based Sm systems.

**FIG. 3 HERE**

However, although all the distributions indicate that the structures will share behaviour at or below ~2 Ti at. %, at higher temperatures, the probability that the expected permutation occurs starts to differ between structures. For example, at 1K the probability the expected permutation occurs is 100% for all structures at ~2 Ti at. %, however, at 300K this drops to 42.6% for SmFe$_{12}$, 38.1% for NdFe$_{12}$, and 15.4% for SmCo$_{12}$. The large disparity between the Fe and Co based systems indicates that having Fe as the transition element makes the structure more predictable.

As these probability distributions are given at each substitution point in a permutation path they can be aggregated across all permutations, to give each positions percentage of the total probability at a given Ti at. % for a range of temperatures. Summing the probabilities of all the positions at each temperature gives a percentage value that indicates how closely the structure adheres to the permutation paths chosen by the simulations search criteria (minimum energy), which acts as a measure of the reliability of its behavior at that temperature. This was used to compare the reliability of the investigated structures at >2 Ti at. %, and a graph of the aggregated total probabilities for each structure against temperature is included in figure 4b.

It can be seen from this graph that as noted previously the Fe based systems retain more of their total probability than the Co based systems at higher temperature, which indicates they will follow the permutation paths more reliably, and hence are more predictable structures. This predictability becomes important when trying to microengineer the structure.

A graph of the percentage change in cohesive energy of these structures against Ti at. % is also given in figure 4a. As can be seen from this graph SmCo$_{12}$ gains a much larger percentage change in energy per Ti substitution, and therefore, if it follows the permutation path exactly, it is likely the most stable of the investigated structures.

**FIG. 4 HERE**

## IV. CONCLUSIONS

As can be seen in figure 4a the average percentage gain in cohesive energy per Ti substitution is -0.45% for SmCo$_{12}$ and -0.31% and -0.25% for NdFe$_{12}$ and SmFe$_{12}$ respectively, indicating that SmCo$_{12}$ is the most stable when substituted with Ti. However, as can be seen from figure 4b, SmCo$_{12}$ has a very low probability of following the permutation path that gives these percentage gains in cohesive energy, even at relatively low temperatures, whereas the Fe based systems retain a greater percentage of their probabilities. Therefore, a hybrid structure using both Fe and Co as the transition metal in the 1:12 phase, may be a prudent way to solve both the need for greater gains in stability per Ti substitution, and the need for a predictable structure.

### ACKNOWLEDGMENT

This work is based on results obtained from the future pioneering program "Development of magnetic material technology for high-efficiency motors" commissioned by the New Energy and Industrial Technology Development Organization (NEDO).

Thank you to the Toyota Motor Corporation for sponsoring this work.

REFERENCES

[1] J.D. Livingston, "The History of Permanent-Magnet Materials.", February 1990, Volume 42, Issue 2, p. 30.
[2] M. Sagawa, S. Hirosawa, H. Yamamoto, S. Fujimura, and Y. Matsuura, "Nd–Fe–B permanent magnet materials.", Jpn. J. Appl. Phys. 26 (1987), p. 785.
[3] Mark Humphries, "Rare Earth Elements the Global Supply Chain.", Congressional Research Service, 2013
[4] D. B. De Mooij, and K. H. J. Buschow, "Some novel ternary ThMn12-type compounds.", JLCM, 136 (1988) pp. 207
[5] Y. C. Yang, X. D. Zhang, L. S. Kong, Q. Pan, and S. L. Ge, "Magnetocrystalline anisotropies of RTiFe11N x compounds.", Appl. Phys. Lett. 58, (1991) p. 2042
[6] S. Hirosawa, Y. Matsuura, H. Yamamoto, S. Fujimura, M. Sagawa, and H. Yamauchi, "Magnetization and magnetic anisotropy of R2Fe14B measured on single crystals.", J. of Appl. Phys. 59, (1986), p. 873
[7] T. Miyake, K. Terakura, Y. Harashima, H. Kino, and S. Ishibashi, "First-principles study of magnetocrystalline anisotropy and magnetization in NdFe12, NdFe11Ti, and NdFe11TiN.", J. Phys. Soc. Jpn. 83 (2014), 043702
[8] J.D. Gale, "GULP: A computer program for the symmetry-adapted simulation of solids", J.C.S. Faraday Transactions, (1997), 93, p. 629
[9] S.C. Westmoreland et al., "Multiscale model approaches to the design of advanced permanent magnets.", Scripta Materialia, 148, (2018) p. 56
[10] P. M. Morse, "Diatomic molecules according to the wave mechanics. II. Vibrational levels.", Phys. Rev. 34, 57, 1929
[11] W. Q. Wang et al. "Structural and magnetic properties of RCo12-xTix (R= Y and Sm) and YFe12-xTix compounds. J. Phys.", D: Appl. Phys. 34 (2001) p. 307
[12] S. J. Blundell, and K. M. Blundell, "Concepts in Thermal Physics", (2006), p. 37

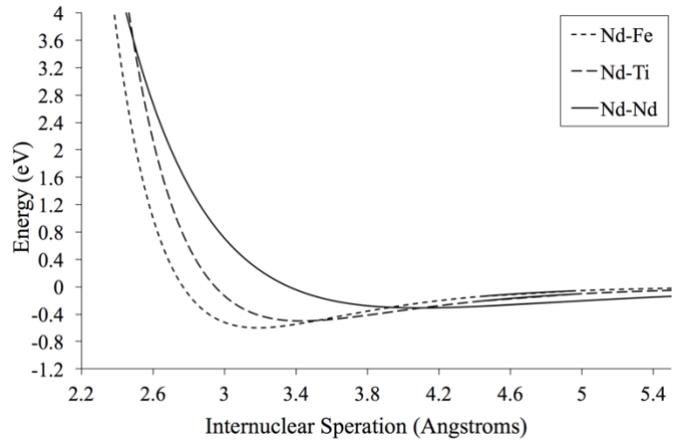

Fig. 1. Interatomic potentials for Nd-Fe, Nd-Ti, and Nd-Nd.

TABLE I
LATTICE CONSTANT COMPARISON

| Structure | Expected | | | Calculated | | |
|---|---|---|---|---|---|---|
| | a | b | c | a | b | c |
| $NdFe_{11}Ti$ | 8.574 | 8.574 | 4.907 | 8.554 | 8.553 | 4.853 |
| $SmFe_{11}Ti$ | 8.557 | 8.557 | 4.800 | 8.482 | 8.494 | 4.814 |
| $SmCo_{11}Ti$ | 8.426 | 8.426 | 4.741 | 8.508 | 8.513 | 4.821 |

Shows a comparison of the expected and calculated values for the various structures lattice parameters. All values are given in Angstrom. The expected values for $NdFe_{11}Ti$ and $SmFe_{11}Ti$ can be found in [5], the expected values of $SmCo_{11}Ti$ can be found in [4].

TABLE II
POSITION UNIT CELL AND FRACTIONAL COORDINATES

| Position | Unit cell | a | b | c |
|---|---|---|---|---|
| 1 | [0,0,0] | 0.1784 | 0 | 0 |
| 2 | [0,0,0] [0,1,0] | 0.1784 | 0.5 | 0 |
| 3 | [1,0,0] | 0.6784 | 0 | 0 |
| 4 | [1,0,0] [1,1,0] | 0.6784 | 0.5 | 0 |
| 5 | [0,0,0] | 0.4284 | 0.25 | 0.5 |
| 6 | [0,1,0] | 0.4284 | 0.75 | 0.5 |
| 7 | [1,0,0] | 0.9284 | 0.25 | 0.5 |
| 8 | [1,1,0] | 0.9284 | 0.75 | 0.5 |
| 9 | [0,0,0] | 0.3216 | 0 | 0 |
| 10 | [0,0,0] [0,1,0] | 0.3216 | 0.5 | 0 |
| 11 | [1,0,0] | 0.8216 | 0 | 0 |
| 12 | [1,0,0] [1,1,0] | 0.8216 | 0.5 | 0 |
| 13 | [0,0,0] | 0.0716 | 0.25 | 0.5 |
| 14 | [0,1,0] | 0.0716 | 0.75 | 0.5 |
| 15 | [1,0,0] | 0.5716 | 0.25 | 0.5 |
| 16 | [1,1,0] | 0.5716 | 0.75 | 0.5 |
| 17 | [0,0,0] | 0 | 0.3216 | 0 |
| 18 | [0,1,0] | 0 | 0.8216 | 0 |
| 19 | [0,0,0] [1,0,0] | 0.5 | 0.3216 | 0 |
| 20 | [0,1,0] [1,1,0] | 0.5 | 0.8216 | 0 |
| 21 | [0,0,0] | 0.25 | 0.0716 | 0.5 |
| 22 | [0,1,0] | 0.25 | 0.5716 | 0.5 |
| 23 | [1,0,0] | 0.75 | 0.0716 | 0.5 |
| 24 | [1,1,0] | 0.75 | 0.5716 | 0.5 |
| 25 | [0,0,0] | 0 | 0.1784 | 0 |
| 26 | [0,1,0] | 0 | 0.6784 | 0 |
| 27 | [0,0,0] | 0.5 | 0.1784 | 0 |
| 28 | [0,1,0] [1,1,0] | 0.5 | 0.6784 | 0 |
| 29 | [0,0,0] | 0.25 | 0.4284 | 0.5 |
| 30 | [0,1,0] | 0.25 | 0.9284 | 0.5 |
| 31 | [1,0,0] | 0.75 | 0.4284 | 0.5 |
| 32 | [1,1,0] | 0.75 | 0.9284 | 0.5 |

Lists all 32 8i positions in a 2x2 supercell of the 1:12 phase structure, the unit cell they occupy, and their fractional coordinates within the supercell.

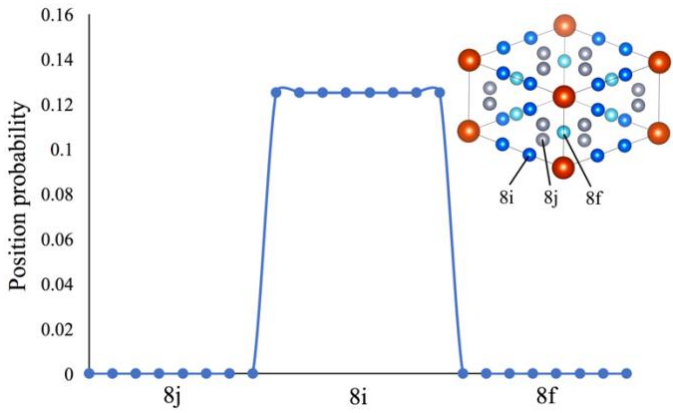

Fig. 2. A graph showing the probability distribution for Ti across the 8i, 8j, and 8f sets of atomic positions at 300K. In the top right hand corner is a singular unit cell of the 1:12 phase structure, the atoms sets are labelled and coloured by position, dark blue is the 8i position set, grey is the 8j position set, and light blue is the 8f position set.

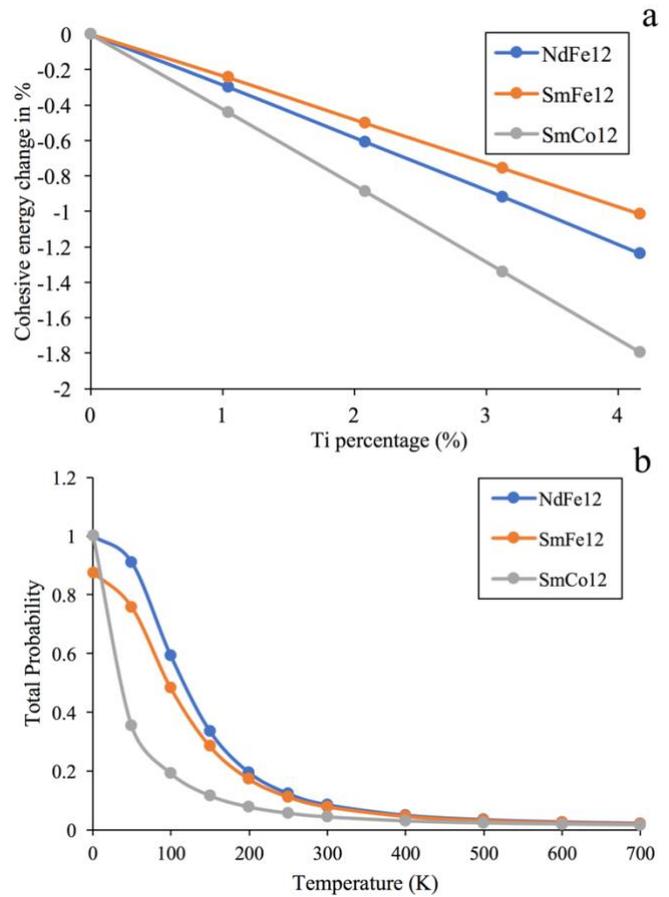

Fig. 4. (a). A graph of the percentage change in cohesive energy against Ti at. % for $NdFe_{12}$, $SmFe_{12}$, and $SmCo_{12}$, (b). A graph of the total probability contained by the minimum energy criteria at the fourth Ti substitution extrapolated from 0K up to 700K for $NdFe_{12}$, $SmFe_{12}$, and $SmCo_{12}$.

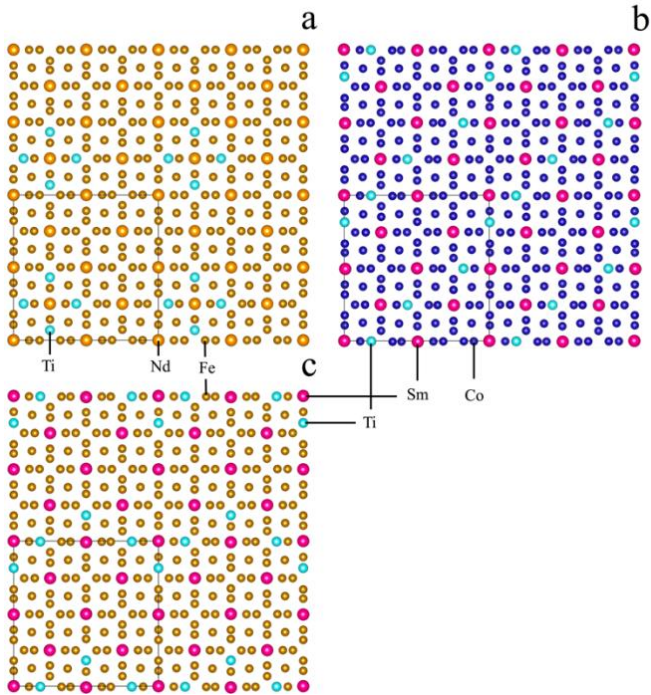

Fig 3. Shows the minimum energy structures of the investigated 2x2 supercells at ~4 Ti at. %, the black box seen in the bottom left corner of each unit cell encloses one supercell structure, (a). $NdFe_{12-x}Ti_x$, (b). $SmCo_{12-x}Ti_x$ in [110] direction, (c). $SmFe_{12-x}Ti_x$ in [100] direction. In all diagrams Ti is light blue, Nd is light orange, Fe is a dark orange, Sm is pink, and Cobalt is dark blue, or as labelled.